\documentclass[12pt,amssymb]{article}

\let\d=\delta

\let\l=\lambda
\let\m=\mu\let\n=\nu
\let\s=\sigma

\let\D=\Delta

\newcommand{\be}{\begin{equation}}
\newcommand{\ee}{\end{equation}}
\newcommand{\bea}{\begin{eqnarray}}
\newcommand{\eea}{\end{eqnarray}}

\newcommand{\del}{\partial}

\begin{document}

\begin{titlepage}
\begin{center}
\vskip .2in \hfill \vbox{
    \halign{#\hfil         \cr
           hep-th/0505054 \cr
           UCI-TR-2005-15\cr
           UCSD-PTH-05-03\cr
           MCTP-05-74 \cr
           May 2005    \cr
           }  
      }   
\vskip 1.5cm {\Large \bf High-Energy Scattering in
Non-Commutative Field Theory  } \\
\vskip .1in \vskip .3in {\bf  Jason Kumar}\footnote{e-mail
address: jpkumar@umich.edu}\\
\vskip .15in {\em Michigan Center for Theoretical Physics,
University of Michigan,\\
Ann Arbor, MI  48105 USA \\} \vskip .2in {\bf Arvind
Rajaraman}\footnote{e-mail address: arajaram@uci.edu} \vskip .15in
{\em Department of Physics,
University of California, Irvine,  \\
Irvine, CA 92697 USA\\} \vskip .1in \vskip 1cm
\end{center}
\begin{abstract}
We analyze high energy scattering for non-commutative field theories
using the dual gravity description.  We find that the
Froissart-Martin bound still holds, but that cross-sections stretch
in the non-commutative directions in a way dependent on the infrared
cutoff.  This puzzling behavior suggests new aspects of UV/IR mixing.
\vskip 0.5cm
\end{abstract}
\end{titlepage}
\newpage

\section{Introduction}

In standard quantum field theory, the cross-section for a process
cannot grow arbitrarily quickly at high energies. This is formalized
by the Froissart-Martin bound \cite{Froissart:ux,Martin:1962rt},
which states that the cross-section can only increase with energy at
most as \bea \sigma = C \log^2 s \eea where $C$ is constant.

This bound is only relevant for theories with no massless particles.
In particular, quantum gravity appears to violate this bound. At high
energies, the cross-section is dominated by black hole production
\cite{Thorne,Eardley:2002re,Banks:1999gd,Giddings:2001bu} and since
the radius of the black hole increases with the energy, one expects
the cross-section to show power-law increase as one goes to the UV.

These competing intuitions come into conflict in the case of
non-commutative field theory (NCFT). Although they are field
theories, they display many properties of quantum gravity in general
and string theory in particular
\cite{Seiberg:1999vs,Minwalla:1999px,VanRaamsdonk:2000rr,Douglas:2001ba,
Fischler:2000fv,Fischler:2000bp,Arcioni:2000bz,Intriligator:2001yu,Armoni:2001uw,
Intriligator:2001ii,Armoni:2002fh}.
Since a NCFT contains interactions of arbitrarily high dimension, the
standard field-theoretic arguments for the Froissart bound do not
necessarily apply.  And although a NCFT can
 contain a naive mass gap, UV/IR mixing
 \cite{Minwalla:1999px,VanRaamsdonk:2000rr}
 implies that
poles at zero momentum may nevertheless arise due to UV effects. As a
result, it seems quite possible that NCFTs, even with a naive mass
gap, may nevertheless violate the Froissart bound and display a
power-law growth more consistent with quantum gravity intution.

In quantum gravity, general arguments indicate that high-energy
scattering is dominated by black hole production.  One might
similarly expect that non-perturbative contributions to scattering in
NCFT become dominant at high energies.  Non-perturbative
contributions are expected to scale as $e^{1\over g^2}$, where $g$ is
a dimensionless coupling of the theory.  In NCFT there are
dimensionless couplings of the form $\theta E^2$, where $\theta$ is
the non-commutativity parameter and $E$ is the characteristic energy
of the process.  Since this theory does not break down at any scale,
it should be valid up to energies at which $\theta E^2 \sim 1$, when
non-perturbative corrections should become important to scattering.
And indeed, NCFT contains non-perturbative solitons
\cite{Gopakumar:2000zd} \cite{Harvey:2000jt} \cite{Ohta} 
whose size scale with
area which would be just right to play the role of the black hole in
scattering. Furthermore, there is a stretched-string effect in NCFT
\cite{Liu:2000qh} whereby the size of bifundamental particles
increases in the non-commutative directions.

All of this makes it plausible that high-energy scattering in NCFT is
dominated by non-commutative solitions, in analogy to the black holes
of quantum gravity, but in contrast to our usual notions of field
theories. Our goal in this paper will be to use the AdS/CFT
correspondence (in a manner analagous to \cite{Giddings:2002cd} in
the commutative context) to test this intuition. This should provide
insight into the nature of the deep relation between NCFT and quantum
gravity.

In section 2, we review the calculation, due to Giddings
\cite{Giddings:2002cd}, of the Froissart bound in commutative field
theory from AdS/CFT.  In section 3 we use an analogous calculation to
determine the scaling of cross-sections with energy in
non-commutative field theory.  In section 4, we find that
cross-sections stretch in the non-commutative directions, and in
section 5 we conclude with a discussion of how these results impact
our understanding of UV/IR mixing.

\section{The Froissart Bound from AdS/CFT}

It was shown by Giddings \cite{Giddings:2002cd}
({and further elucidated by Kang and Nastase
\cite{Kang:2004jd,Kang:2005bj}}) that the Froissart-Martin bound for
a gauge theory could be derived using the AdS/CFT correspondence. As
is well-known, string theory in an $AdS_5 \times S^5$ background is
dual to $N=4$ gauge theory on the boundary of spacetime. According to
AdS/CFT, scattering amplitudes in the boundary theory correspond to
scattering amplitudes in the bulk theory. The corresponding
cross-section for the bulk scattering process, when projected onto
the brane, yields the cross-section for the gauge theory process
\cite{Polchinski:2001tt}\cite{Boschi-Filho:2002zs}. In
fact, high energy bulk gravitational scattering is dominated by black
hole production, and the corresponding behavior of the bulk
cross-section precisely corresponds  to a process which saturates the
Froissart-Martin bound in the dual boundary theory.

In the boundary gauge theory, the momentum is \be p_{a} =-\imath
{\partial \over \partial x^{a}} \ee This momentum is a conserved
quantity, as it generates an isometry (translation by $x^{a}$). But
an observer in the bulk will measure a momentum \be \tilde p_{\mu} =
e_{\mu} ^{~a} p_{a} \ee in a local inertial frame. Bulk amplitudes
will depend on this momentum evaluated in the bulk at the point of
interaction.  An incoming boundary scattering state will appear in
the bulk scattering process as an incoming wavefunction of the form
\be \psi \sim e^{ip\cdot x} f(u) \ee where we ignore the dependence
on the transverse $S^5$.  The bulk and boundary scattering amplitudes
are then related by \be A_{gauge} (p) =\int d^5 x \sqrt{-g}
A_{bulk}(x^{\mu},u) \Pi_i \psi_i \ee

We can find the bulk cross section by assuming that the high energy
behavior is dominated by black hole formation. The mass of the black
hole will be the bulk energy of the collision. The cross-section for
bulk scattering is then proportional to the square of the black hole
radius. Having computed the bulk cross-sectional radius $\tilde r$,
we can determine the scale $r$ of the corresponding gauge theory
process by inverting the relation: \be \tilde r_{\mu} = e_{\mu} ^{~a}
r_{a} \ee

We can now derive the Froissart bound for ordinary gauge theories,
following \cite{Giddings:2002cd}. We start with the $AdS_5\times S^5$
metric \bea
ds^2=\alpha'R^2\left[u^2(-dx_0^2+dx_1^2+dx_2^2+dx_3^2)+{du^2\over
u^2}+d\Omega_5^2\right]
\nonumber\\
=\alpha'\left[dy^2 +e^{-2y/R}dx^{\mu}
dx_{\mu}+R^2d\Omega_5^2\right]~~~~~~~~~~~ \eea
 where $\mu$ runs from 0 to 3, and $u={1\over R}e^{-{y\over R}}$.
 The boundary is at ${1\over u}=z=0$.

The interpretation of scattering in pure $AdS_5\times S^5$ is
problematic because the maximally supersymmetric gauge theory is
conformal. This can be rectified by considering a deformation of the
gauge theory which makes it non-conformal. Correspondingly, the dual
geometry will be modified in the infrared.

This deformation can be done in many ways; for instance one could
consider the $N=1^*$ theory of \cite{Polchinski:2000uf}, or the
finite temperature theory \cite{Witten:1998zw}. Since the high-energy
behavior is expected to be universal, we will simply truncate the
geometry by putting an infrared cutoff brane at $u=u_c$
\cite{Giddings:2002cd}. Without loss of generality, we can take
$u_c=R^{-1}$.

The high energy behavior of scattering processes is dominated by the
production of black holes in this geometry. Unfortunately, the exact
metric of these black holes is unknown. We can however estimate the
size of the black holes by using a procedure suggested by Giddings.
In this procedure, one solves for the linearized perturbations
produced by a mass $m$ in this geometry. The location where the
perturbations become of order 1 should approximately reproduce the
horizon geometry.

For this purpose, we can rewrite the metric as \bea ds^2
=(1+h_{yy})dy^2 +e^{-2y/R}(\eta_{\mu \nu} +h_{\mu \nu}) dx^{\mu}
dx^{\nu} \eea where $h$ is the perturbation of the metric (we are
ignoring any dependence on the $S^5$, since we are looking for
spherically symmetric solutions.)

In an appropriate gauge, the linearized metric fluctuations will
satisfy the bulk equation
\cite{Giddings:2002cd}\cite{Maldacena:1999mh} \bea D^2 h_{\m\n}=0
\eea

We need to supplement this with boundary conditions in the UV and IR.
In the UV ($z\sim 0$), the condition is that the field needs to fall
off sufficiently rapidly. In the IR, we cutoff the space with a brane
which has some fixed energy density $T_{\m\n}=S_{\m\n}\d(y)$.  The
boundary conditions for fields at this brane may be determined by
integrating the equations of motion over a small region around the
brane.  We consider the approximation in which the brane is rigid
(the so-called ``heavy radion" limit) and find the boundary condition
\bea \del_y(h_{\m\n}-\eta_{\m\n}h)|_{y=0} -{d-1\over R} h_{yy}(0)
\eta_{\mu \nu } ={S_{\m\n}\over 2M_p^{d-1}} \eea where $h=\eta^{\mu
\nu} h_{\mu \nu}$, $u={1\over R}e^{-{y\over R}}$ and $d=4$. The
linearized equation of motion for $h$ then is solved by: \bea
h_{\m\n} = {1\over 2M_p^{d-1}}\int {d^d x'\over (2\pi)^d}\sqrt{g}
\Delta_{d+1} (X,0;x')\times
\nonumber\\
\left[ S_{\m\n}(x')-\eta_{\m\n}{S_{\rho}^{~\rho}(x') \over
d-1}+{\del_{\m}\del_{\n} \over \del^2}{S_{\rho}^{~\rho}(x') \over
d-1}\right] \eea where the function $\D$ satisfies \bea
D^2\Delta_{d+1} (X,0;x')={\d^{d+1}(X-X')\over \sqrt{G}}
\\
\del_y\Delta_{d+1} (X,X')|_{y=0}=0 \eea $\Delta$ is thus the Neumann
Green's function for the scalar Laplacian \bea \Delta_{d+1}
(x,u,x',R)=-\left({1\over uR}\right)^{d\over2}\int {d^dp\over
(2\pi)^d}{1\over q}{J_{d\over 2}({q\over u})\over J_{{d\over
2}-1}(qR)}e^{ip(x-x')} \eea with $q^2 =-p^2$.

If we place a mass $m$ at the point $u^{-1}=R,x'=0$, we will produce
perturbations around the background metric. We then estimate the
horizon of the black hole as the place where the perturbations are
${\cal O}(1)$: \bea 1\sim {m\over 2M_p^{d-1}} \sqrt{g} \Delta_{d+1}
(X,0;0) \eea

One can compute the $p$ integral by contour integration, and the
integral will pick up the poles of the integrand at the zeros of the
denominator. We are only interested in the limit of large $m$ (where
$m$ is the mass of the black hole) and this limit is dominated by
$q\sim 0$, leaving

\bea \Delta_{d+1} (x,u,x',R)=
   \int {d^{d}p\over (2\pi)^d}{1\over q}{({qR\over
2})^{{d\over 2}} \over J_{{d\over 2}-1}(qR)} {1\over \Gamma({d\over
2}+1)} e^{ip(x-x')} \eea

The integral is dominated by the first zero of the denominator,
$q=M_1$. Evaluating the residue at this pole and performing the
angular integrations, one finds that the linearized perturbation
satisfies \bea h_{00} &\propto& m u^{-4} {e^{-M_1 r} \over r} \eea
where $r$ is the distance in the longitudinal directions, the UV
boundary is at $u=\infty$ ($y=-\infty$) and the IR brane is at
${1\over u}=R$ ($y=0$). The pole in $\Delta_p$ is located at
$p=\imath M_1$. Setting $h_{00} \sim 1$ and taking the logarithm of
both sides gives \bea M_1 r +\log(M_1 r) \sim \log(mM_1) + 4y \eea
Near $r=0$ the solution is expected to be smoothed out
\cite{Giddings:2002cd}. Since the warp factor must be monotonic, the
horizon size is maximized at the IR brane ($y=0$) and decreases as
$y$ decreases (i.e. as one moves to the UV).  Thus the longitudinal
size of the horizon is maximized when \bea r \sim {1\over M_1} \log
({m\over M_1}) \eea This corresponds to a cross section $\s=\pi
r^2\sim (\ln s)^2$. The cross-section hence saturates the Froissart
bound.

\section{The Noncommutative Case}

We will now use similar arguments to find the Froissart bound for the
case of noncommutative field theory. The gravity solution dual to an
NCFT was found in \cite{Maldacena:1999mh} and
\cite{Hashimoto:1999ut}. This background is described by the solution
\bea \label{ncback}
ds^2=\alpha'R^2\left[u^2(-dx_0^2+dx_1^2)+{u^2\over
1+a^4u^4}(dx_2^2+dx_3^2))+{du^2\over u^2}+d\Omega_5^2\right]
\nonumber\\
B_{23}=\alpha'R^2 {a^2u^4\over 1+a^4u^4} \qquad\qquad\qquad
e^{2\varphi}={g^2\over 1+a^4u^4}~~~~~~~~~~~~
\nonumber\\
A_{01}=\alpha'{b\over g}u^4R^4 \qquad\qquad\qquad
F_{0123u}={{\alpha'}^2\over g(1+a^4u^4)}\del_u(u^4R^4)\eea

The boundary of the space is $u=\infty$. For small $u$, the solution
becomes $AdS_5\times S^5$ (with the relation $u={1\over z}$).

We shall now put a cutoff brane at $u=u_c$. This will correspond to
the cutoff brane considered by \cite{Giddings:2002cd}. Scattering
processes at high energy will then correspond in the dual theory to
the formation of a black hole of mass $M$ at the point $u=u_c$. To
find the approximate shape of the black hole, we shall consider the
linearized response to a source placed at the origin of $x_i$ and at
$u=u_c$. The horizon will be approximately at the place where the
fluctuations become large and the linearized approximation breaks
down.

The calculation in this background will be a good deal more
complicated than the calculation in $AdS^5\times S^5$, because the
background geometry is less symmetric and involves more fields.
Perturbations of the metric will now be coupled to perturbations of
the gauge fields. Furthermore, in this case, unlike $AdS^5$, there is
a new scale $a$ in the problem. The shape of the black hole will now
depend on the relative sizes of $u_c$ and $a$. To avoid these
technical complications, we shall instead consider as a model the
fluctuations of a scalar field with the action \bea S=\int d^{10} x
\sqrt{g}e^{-2\varphi}(\del\phi)^2\eea and Neumann boundary
conditions. (Here $\varphi$ is the background dilaton, and $\phi$ is
a different field.)

 The equation of motion for the field is \bea
\del_\mu(\sqrt{g}g^{\m\n}e^{-2\varphi}\del_\n \phi)=0 \eea For
linearized fluctuations around the geometry (\ref{ncback}), this
reduces to \bea {1\over u^{5}} \del_u(u^5\del_u\phi)+{1\over
u^{4}}(-\del_0^2+\del_1^2)\phi+{1\over u^{4}}(1+a^4u^4)
(\del_2^2+\del_3^2)\phi=0 \eea

We shall further assume that the scalar is coupled to the external
source mass $M$ by a coupling \bea S=\int d^{10} x \sqrt{g}
Me^{-2\varphi}\phi
\eea Then in the presence of a mass $M$ located at $u=u_M,
x_1=x_2=x_3=0$, the equation of motion is modified to be \bea
\label{nceomsource} {1\over u^{5}} \del_u(u^5\del_u\phi)-{1\over
u^{4}}k_1^2\phi-{1\over u^{4}}(1+a^4u^4) (k_2^2+k_3^2)\phi=M{1\over
u^{5}}\d(u-u_M) \eea where we have Fourier transformed $\phi$ by \bea
\phi(x,u)=
  \int d^3k\
e^{ik_1x^1+ik_2x^2+ik_3x^3}\phi(k_1,k_2,k_3;u) \eea

We will initially place the black hole source in the bulk of
space-time, and then take the limit as it approaches the IR cutoff
brane (i.e. $u_M\rightarrow u_c$).  We will designate the solution
for $\phi$ in the region between the black hole and the IR cutoff as
$\phi_-$, and the solution for $\phi$ in the region between the black
hole and the UV cutoff as $\phi_+$,

Regularity of the solution mandates that $\phi_+$ should fall off as
$u\rightarrow \infty$.
 On the other hand, $\phi_-$
 satisfies a
 Neumann boundary
condition at the brane $\del_u\phi_-=0$ at $u=u_c$. $\phi_+$ and
$\phi_-$ must be matched by integrating (\ref{nceomsource}) across
the source.

In the limit $u_M\rightarrow u_c$ we can focus on $\phi_+$. The
entire solution is then determined to be \bea \phi(x,u)={M\over
u_c^5}
  \int d^3k\
e^{ik_1x^1+ik_2x^2+ik_3x^3} {\phi_+(k_1,k_2,k_3,a;u)\over
\del_u\phi_+(k_1,k_2,k_3,a;u=u_c)
 } \eea

When we look for the limiting cross-section, we want to maximize the
black hole area; that is, we want to know the largest radius where
$\phi$ becomes of order 1. This must occur at $u=u_M$, due to the
monotonicity of the warp factor. Hence we must solve the equation
\bea {M\over u_c^5}
  \int d^3k\ e^{ik_1x^1+ik_2x^2+ik_3x^3}
{\phi_+(k_1,k_2,k_3,a;u_c)\over \del_u\phi_+(k_1,k_2,k_3,a;u_c)
  } =1 \eea

To get an idea of the shape of the black hole, we will look for an
solution for $\phi_+$. The exact solution is unwieldly and not very
helpful, so we instead will look for an approximate solution. We
shall do this by solving the equation (\ref{nceomsource}) for $u\gg
{1\over a}$ and $u\ll {1\over a}$ separately, and matching them at
$u\sim a$.

First, we consider the case $u\gg {1\over a}$. Here the equation for
$\phi_+$ simplifies to \bea {1\over u^{5}} \del_u(u^5\del_u\phi)
-a^4(k_2^2+k_3^2)\phi=0 \eea with the solution \bea \phi=u^{-
2}\left(C_1K_\nu(\l u)+C_2I_\nu(\l u)\right) \eea where $\nu=\pm 2$
and $\l^2=a^4(k_2^2+k_3^2)$.

For large $u$ the UV boundary condition requires $C_2=0$. Therefore,
in this region \bea \phi=u^{- 2}C_1K_2(\l u) \eea

For $u\ll {1\over a}$ the equation simplifies to \bea {1\over u^5}
\partial_u u^5
\partial_u \phi -p^2 ({1\over u^4}) \phi &=& 0
\eea where we have defined $p^2 =k_1^2 +k_2^2+k_3^2$.
 The solution is found to be
\bea \phi=u^{-2}(C_5K_{2}(p u^{-1})+C_6I_{2}(p u^{-1})) \eea

The near and far solutions should match in the intermediate region
$u\sim a$. This then imposes the relations
 \bea
 C_5=0\qquad C_6
 = 16 C_1p^{-2}\l^{-2}
  \eea

The approximate scalar field profile is then
 \bea
\  \phi_+=\left\{ \begin{array}{ll} 16  u^{-2}I_{2}(p u^{-1}) &{u\ll a^{-1}}\\
u^{- 2}p^2\l^{2}K_2(\l u) &{u\gg a^{-1}}
\end{array}\right.
 \eea
 We have defined $p^2=k_1^2 +k_2^2+k_3^2$, $\l^2=a^4(k_2^2+k_3^2)$.

 Note that the solution for ${u\ll a^{-1}}$ is
 symmetric in all the directions, as it should be, since the background solution
 approaches the $AdS_5$ geometry in this limit. On the other hand, the
  solution for ${u\gg a^{-1}}$ is not spherically symmetric.

Finally, we can now use these approximate solutions to find the shape
of the resulting black holes. Let us first  assume {$a u_c \ll 1$}.
We should then use the form of $\phi_+$ for  ${u\ll a^{-1}}$. The
integral equation is then \bea {M\over u_c^5}
  \int d^3k\  e^{ik_1x^1+ik_2x^2+ik_3x^3} {  u_c^{-2}I_{2}(p
u_c^{-1})\over
  \del_u( u^{-2}I_{2}(p
u^{-1}))|_{u=u_c}} =1 \eea This is exactly the same integral that one
obtains for the $AdS$ case. This is to be expected, since the
geometry looks like $AdS$ for $a u_c \ll 1$. As in that case, the
integral can be evaluated by contour integration and is dominated by
the poles in the denominator.  The dominant pole will appear at
$p=\imath M_1$, where $M_1$ is determined by the precise form of the
denominator (the pole must be imaginary, because the graviton
solution is real for real momenta). One then finds $h_{00} \propto
{e^{-M_1 r} \over r}$.  We therefore get the same results as before
(i.e., $\sigma \sim {1\over m_{\pi}^2} \ln^2 {E\over m_{\pi}^2}$).

More concretely, we can expand the numerator near zero to get \bea
e^{-\sqrt{u_c^2-k^2}x^i} e^{-kx}Mu^{- 2}R^2a^4 k^2(a^2k u_c)^{-2} =
e^{-\sqrt{u_c^2-k^2}x^i} e^{-kx}MR^2( u_c)^{-4} \eea The numerator is
spherically symmetric, so we get a spherical black hole whose size is
independent of $a$. This is to be expected; for small $a$, the
effects of noncommutativity are pushed out to very large $u$, so the
black hole doesn't feel the noncommutativity. It doesn't matter how
large the black hole gets, since that only modifies the shape far
out, not near the brane.

Now take {$au_c \gg 1$}.  We should then use the form of $\phi_+$ for
${u\gg a^{-1}}$. The integral equation is then \bea {M\over u_c^5}
  \int d^3k\  e^{ik_1x^1+ik_2x^2+ik_3x^3}
  {u_c^{- 2}K_2(\l u_c)\over \del_u( u^{- 2}K_2(\l u))|_{u=u_c}}
=1 \eea

The integral over $k_1$ is now trivial (since $\l^2\equiv
a^4(k_2^2+k_3^2)$), and we get a $\d(x_1)$ factor. The remaining
integral can be performed as before (by contour integration, which is
dominated by the pole), and we find a logarithmic growth in the $x_2,
x_3$ directions.

We expect the delta function to be smoothed out by subleading terms.
Indeed, this form of the far solution is not valid when $k_1$ is
large.\footnote{equivalently, it is valid if ${r_2 ^2+r_3 ^2 \over
r_1 ^2} \ll (au)^4$} If the integrand has support for large $k_1$,
then this corresponds to a small but non-zero size for the black hole
in the $x_1$ direction.

 Note that the precise form
of the denominator, as in the original $AdS_5$ case, is not very
important. This is natural, as the Froissart bound should be a very
general result.  Indeed, we have only implemented the IR cutoff in a
heuristic manner, modeled by cutting off the space with an IR brane.
The fact that we can nevertheless see the Froissart-Martin bound in
the $AdS_5$ case is an indication of how resilient this result is to
changes in boundary conditions.

For this simple toy model, we thus get two results: if the infrared
cutoff is taken to be much smaller than ${1\over a}$, the shape of
the black hole is very similar to the pure $AdS$ case. On the other
hand, if the infrared cutoff is taken to be very large, the shape of
the black hole is very different -- it is much smaller in the
commutative directions, and still grows logarithmically in the
noncommutative directions.

\section{Scaling in the Commutative Directions}

Our analysis of the scalar fluctuations has determined the parametric
dependence of the size of the cross-section in the non-commutative
directions.  We have found that the size in the commutative
directions is much smaller by comparison (when $au_c \gg 1$), but we
would like to determine its behavior more precisely.

Now, if the cutoff $u_c$ satisfies $u_ca\ll 1$, we should look at the
behavior of $\psi_+$ near $u\sim 0$. In this region, the background
geometry reduces to the $AdS_5\times S^5$ geometry, and the behavior
of the fluctuations will be the same as the $AdS$ behavior. So the
shape of the black hole in this limit will be the same as the shape
found in \cite{Giddings:2002cd}.

The interesting case is therefore $u_ca\gg 1$, where the
noncommutativity is important. In this region, the geometry again
simplifies. In particular, the metric becomes \bea
ds^2=\alpha'R^2\left[u^2(-dx_0^2+dx_1^2)+{1\over
a^4u^2}(dx_2^2+dx_3^2)+{du^2\over u^2}+d\Omega_5^2\right] \eea

Now define $u=u_ce^\s, \tilde{x}_0=u_cx_0, \tilde{x}_1=u_cx_1,
 \tilde{x}_2={1\over
a^2u_c}x_2, \tilde{x}_3={1\over a^2u_c}x_3$. Then the metric becomes
\bea ds^2=\alpha'R^2\left[e^{2\s}(-d\tilde{x}_0^2+d\tilde{x}_1^2)+
 e^{-2\s}(d\tilde{x}_2^2+d\tilde{x}_3^2)+{d\s^2}+d\Omega_5^2\right]
\eea independent of $a, u_c$.
 The IR cutoff brane (where the source is located) is at
$\sigma =0$, which is also independent of $a, u_c$.

In this geometry, all the coefficients are of order 1, so we expect
the size of the black hole to be parametrically the same in all
directions. Furthermore, the calculations of the previous sections go
through, and we find that the size should scale as $\ln m$ in the
non-commutative directions.  Thus, they should should scale as $\ln
m$ in all the directions.

Rescaling back to the original coordinates, we find that the black
hole radii should be \bea r_1\sim {\ln m\over u_c}\qquad r_2\sim
r_3\sim {(\ln m) a^2 u_c} \eea which is consistent with our earlier
result for the relative size of the cross-section in commutative and
non-commutative directions.

\section{Discussion}

We have found that the scattering cross-section in NCFTs still obeys
a Froissart-like bound $\s\propto (\ln s)^2$ at high energies. The
interesting new feature is that the growth is not spherically
symmetric; the size of the cross-section in the commutative
directions grows more slowly than the size in the non-commutative
directions\footnote{Related work is the calculation of quantum 
commutators\cite{Alvarez-Gaume:2001ka}, from which one can also 
study fall-off behavior.}.  Furthermore, this ratio of
cross-sections is universal, \bea {r_{2,3}\over r_1} \sim (au_c )^2
\eea and depends only on the non-commutativity parameter $a$ and the
infrared cutoff $u_c$. Most importantly, this aspect ratio is
independent of the energy of the scattering process.

This provides an answer to the question raised in the introduction:
NCFT scattering at high energies is {\it not} dominated by the
production of non-commutative solitons (if it were, then we would
instead expect ${r_{2,3}\over r_1 } \sim E$). In this regard at
least, NCFTs lack a crucial feature of quantum gravity.  Note that
the absence of linear scaling in the non-commutative directions
also indicates that the behavior of the cross-section is not dominated
by the stretched string effect.  One might wonder if this effect is
somehow limited in the full non-perturbative theory at very high
energies.\footnote{We thank R. McNees for a discussion on this
point.}

The size of $r_1$ is indeed as expected.  If we equate $u_c$ with the
mass gap $m_{\pi}$, then the scaling $r_1 \sim {1\over m_{\pi}}
\ln({s\over m_{\pi}^2})$ is the usual Froissart bound. The
coefficient ${1\over m_{\pi}}$ is heuristically related to idea that
at the largest distance at which two particles interact, the
potential energy between them should be approximately the mass of one
quantum of the particle mediating the effective interaction.

For the $r_{2,3}$ directions it seems that the effective mass gap is
given by $m_{\pi'}={1\over a^2 u_c}$, which will be much smaller than
$m_{\pi}$ in the limit $au_c \gg 1$.  Indeed, one might have expected
that the effective mass gap in the non-commutative directions should
be smaller because of the additional poles of the form $p \theta^2 p$
which arise in NCFT from UV/IR mixing.  However, the presence of
poles at zero momentum would seem to suggest that the effective mass
gap should be zero, and Froissart behavior should not appear at all.
Instead, it appears that the strong coupling dynamics of the theory
serves to ``soften" the new poles, leaving a small but nonzero
effective mass gap.  The fact that the effective mass gap scales
inversely with the putative mass gap may be an intriguing signature
of the effective mass gap's origin in UV/IR mixing.

It may also be possible to interpret this scaling as the effect of a
renormalization group flow. In this interpretation, the couplings at
high energies are different in the commutative and noncommutative
directions, and flow in the infrared to the usual NCFT action. The
ratio of the  values of the coupling constants for the two directions
is given by $u_ca$. It would be interesting to see if this can be
made more precise.  Recent work \cite{Hubeny:2005qu} has also
emphasized the non-intuitive behavior of NCFT's in the
non-commutative directions.  It would also be very interesting to
understand the connection between our result and the reduction in
the number of degrees of freedom in the high temperature limit of
NCFT's found in 
\cite{Fischler:2000fv,Fischler:2000bp,Arcioni:2000bz,Cai:1999aw}.

Our work here can be generalized to other cases. Noncommutative
theories appear on the worldvolume of D-branes in curved spaces with
background $H$-fields \cite{Alekseev:1999bs}. It would be interesting
to see if the dual supergravity solution for these branes
\cite{Rajaraman:2002vf} exhibits a behavior similar to that found
here, which would confirm that the Froissart behavior is universal.

Finally, we have used the scalar field as a toy model for the fully
coupled perturbations of the metric and gauge fields.  It would be
very interesting (though extremely difficult) to verify our
qualitative picture with a complete non-linear solution to the fully
coupled equations.

\vskip .3in

{\bf Acknowledgments}

We are grateful to C. Bauer, S. Cullen, J. Davis, M. Douglas, F.
Larsen, B. Grinstein, R. Kaiser, J. Kuti, A. Manohar, R. McNees and
especially S. Giddings, K. Intriligator and H. Nastase for useful
discussions. J. K. would like to thank the University of California,
San Diego, where work on this project was initiated. J. K. is supported
by the Michigan Center for Theoretical Physics and the Department of
Energy.  A. R. is supported
in part by NSF Grant 0354993.

\end{document}